\documentclass[aps,prl,twocolumn,english,superscriptaddress,citeautoscript,preprintnumbers,amsmath,amssymb,floatfix,footinbib]{revtex4-1}
\usepackage{hyperref}
\hypersetup{colorlinks=true,linkcolor=red,citecolor=blue}
\usepackage{amsmath}
\usepackage{graphicx}
\usepackage{dcolumn}
\usepackage{float}

\begin{document}
\title {Electric field control of photoinduced effect in La$_{0.7}$Sr$_{0.3}$MnO$_3$/LaTiO$_3$/SrTiO$_3$ heterostructure}
\author{Pramod Ghising}
\affiliation{Department of Physics, Indian Institute of Technology, Kanpur 208016, India}
\author{Z. Hossain}
\email{zakir@iitk.ac.in}
\affiliation{Department of Physics, Indian Institute of Technology, Kanpur 208016, India}
\begin{abstract}	
In this work we have studied the effect of light and electric field on the electrical properties of La$_{0.7}$Sr$_{0.3}$MnO$_3$ (LSMO) film at 300 K. Taking advantage of it's charge, spin, orbital and lattice degrees of freedom, we have successfully shown that light and electric field together can be used to tune the resistance states of the manganite system. Using applied gate voltage (V$_g$) and light we were able to obtain a change in resistance of about $\pm$ 7.5$\%$. Furthermore, incorporating an ultra thin interfacial LaTiO$_3$ (LTO) layer provides valuable insight on the origin of the photoinduced effect. The observed photoinduced effect is attributed to photoexcited down spin e$_{g\downarrow}$ electrons, interfacial charge injection and manipulation of orbital occupancy depending on the gate voltage and wavelength of light.

\end{abstract}

\maketitle

In the present age of ever evolving technology, the field of spintronics grapples to adapt with the need for efficient devices. The need of the hour are materials with multifunctional properties. Complex oxides which play host to a range of exciting phenomena viz. high temperature superconductivity\cite{1993-Nat-Schilling}, ferromagnetism\cite{1950-Physica-Jonker}, multiferroicity\cite{2007-NatMat-Ramesh} and ferroelectricity\cite{2017-ApRev-Acosta} have taken centerstage as the future of device and technology. Ever since the discovery of high temperature superconductivity \cite{1986-ZPhysB-Bednorz}, the capabilities of these oxides as an alternative for the present day semiconductor based devices have been intensely explored. Now with the inception of various thin film deposition techniques, and the ability to fabricate heterostructures with atomically sharp interfaces, the future of oxide based electronics seems imminent.

In this context, doped manganites have attracted a lot of attention. With it's exotic phenomena like colossal magnetoresistance (CMR)\cite{1993-PRL-Helmolt,1994-Science-Jin}, ferromagnetism (FM)\cite{1950-Physica-Jonker}, half-metallicity \cite {1998-Nature-Park} etc., it presents wonderful opportunity for future device and applications. In La$_{1-x}$Sr$_x$MnO$_3$ (LSMO), the amount of Sr doping leads to a rich phase diagram; and can be used to tune it to various magnetic and electrical states. Crystal field and Jahn-Teller (JT) effect splits the Mn \textit{d} orbital into non-degenerate e$_g$ (d$_{x^2-y^2}$, d$_{3z^2-r^2}$) and t$_{2g}$ (d$_{xy}$, d$_{yz}$ and d$_{xz}$) orbitals \cite{2003-JPhysD_ApplPhys-Haghiri}. Strong Hund's coupling (J= 2 eV) between itinerant e$_g$ and localized t$_{2g}$ electrons lead to parallel alignment of the Mn spins \cite{1997-PRB-Y_Morimoto}. Hopping of e$_g$ electrons between parallel aligned Mn$^{3+}$ and Mn$^{4+}$ ions via the double exchange (DE) mechanism gives rise to its ferromagnetic and metallic properties \cite{1951-PhysRev-Zener}. 

Strong interplay between its lattice, charge, spin and orbital degrees of freedom leads to greater control over its various physical properties. Magnetic exchange interaction is very sensitive to e$_g$ orbital occupancy, which can be tuned between d$_{x^2-y^2}$ and d$_{3z^2-r^2}$. In thin films, this freedom can be used to a great advantage. Using strain one can manipulate the e$_g$ orbital occupancy (between d$_{x^2-y^2}$ and d$_{3z^2-r^2}$) to control the magnetic exchange interaction \cite{1999-JPSJ-Konishi,2010-PRL-Sadoc}. Electric field can also be used to tune the magnetic and electrical properties of manganites \cite{2015-AdvFuncMater-Cui,2009-PRL-Dhoot,2014-NatComm-Cuellar}. Sensitivity of magnetic exchange to orbital occupancy, also makes light as another important tuning parameter. Various photoinduced studies in manganites have shown fascinating physics spanning from persistent photoconductivity \cite{2001-PRB-Cauro,2001-JAP-Dai}, modulation of metal insulator transition \cite{2008-PRL-Takubo}, control of exchange bias \cite{2015-NRS-Sung} to formation of charge ordered clusters \cite{2001-PRB-Liu}. Due to the strong exchange interaction, photoexcited electrons have been known to alter the spin alignments in manganites, causing spin disorder in the system, widely known as the photoinduced demagnetization effect \cite{1998-PRB(R)_Matsuda,2000-JpnJAP-Liu}. Thus the use of light and electric field can be used to tune both spin and charge degrees of freedom. Here we have studied the effect of light and gate voltage on the electrical properties of La$_{0.7}$Sr$_{0.3}$MnO$_3$ thin film. Introducing an ultra thin interfacial layer of LaTiO$_3$ helps to identify the mechanism behind the observed photoinduced effect.  We have observed photoinduced increase in resistance, which can be tuned  with the help of gate voltage. Our observations reveal a complex interplay between optical transitions, carrier injection and orbital occupancy.

Pulsed laser ablation was used to deposit single and bi-layer films of LSMO and LSMO/LaTiO$_3$ (LTO) on (001) oriented SrTiO$_3$ (STO) substrates. Prior to film deposition STO was etched with NH$_4$HF solution to obtain TiO$_2$ terminated surface. For film deposition, nanosecond pulses from an excimer laser (KrF, $\lambda$=248 nm) was fired at the target material with a repetition rate of 2 Hz. Oxygen pressure during film deposition was maintained at 0.2 mbar and 1 $\times$ 10$^{-4}$ mbar for LSMO and LTO, respectively. X-ray diffraction with Cu-K$_{\alpha1}$ radiation was used to confirm epitaxial growth of the films. Electrical transport measurement was done in a Quantum Design physical properties measurement system (PPMS). Transient current measurements were done using Keithly 237 sourcemeter. Photoresistivity measurements were carried out in a 4-probe geometry with gate voltage (V$_g$) applied at the back of the STO substrate (Fig. 2 (d)). V$_g$ was applied throughout the entire measurement, starting from t=0 sec for all samples. Blue (441 nm) and UV (325 nm) line of He-Cd continuous wave laser with power density 680 mW cm$^{-2}$ and 690 mW cm$^{-2}$, respectively, were used for photoillumination. All the photoinduced experiments were carried out at 300 K.

X-ray diffraction (XRD) measurement performed on the films show epitaxial growth, with only (00\textit{l}) oriented film peaks. Fig. 1(a) shows the XRD plot around (002) reflection for LSMO(15nm)/STO and LSMO(15nm)/LTO(10u.c)/STO (henceforth will be known as sample d$_0$ and d$_{10}$, respectively, unless otherwise stated). Both the samples exhibit  metallic behaviour down to 3 K as shown in Fig. 1(b). The inset in Fig. 1(b) shows the metal insulator transition (MIT) in the resistivity curves for samples d$_0$ and d$_{10}$. From the MIT values, the Curie temperature (T$_c$) of the LSMO films are estimated to be greater than 330 K.
\begin{figure}[htp]
	\centering
	\includegraphics[width=1.0\linewidth]{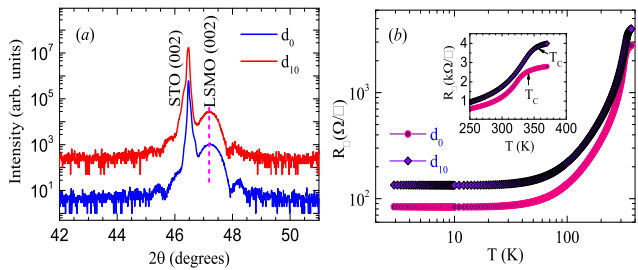}	
	\caption{(Color online) Structural and electrical characterization of sample d$_0$ and d$_{10}$. (a) XRD plot showing (002) reflection of the film and the substrate. (b) Plot of sheet resistivity for Sample d$_0$ and d$_{10}$. Both the samples are metallic down to 3 K. Inset in Fig (b) shows the metal-insulator transition (MIT) in the two samples. From MIT the estimated T$_c$ for both samples are greater than 330 K. }
\end{figure}
The effect of photoexposure on resistivity of LSMO was studied at two wavelengths: UV ($\lambda$=325 nm, E$_{uv}$=3.8 eV) and blue ($\lambda$=441 nm, E$_{blue}$=2.8 eV) at 300 K. In manganites, due to strong spin correlation between the Mn ions, spin of the photoexcited electrons have a strong influence on its magnetic and electrical properties. For LSMO, E$_{uv}$ and E$_{blue}$ correspond to photoinduced down spin (electron) transitions from O2p to Mn e$_{g\downarrow}$ and between e$_g$ orbitals on different Mn sites, respectively \cite{2000-JpnJAP-Liu,1998-PRB-Quijada}. 

In Fig. 2(a)-(c) we show the $\%$ resistance change on exposure to UV and blue laser for sample d$_0$ at 300 K. An increase in resistance (or photoresistivity (PR)) on exposure to UV and blue light is observed at V$_g$=0 (Fig. 2(a)). Resistance change is defined as, ${\Delta}R=R(L)-R(D)$, where ${R(D)}$ and ${R(L}$) are resistance of the sample before and after illumination, respectively. At T $<$ T$_C$, strong Hund's coupling ensures parallel spin alignment between itinerant e$_g$ and localized t$_{2g}$ electrons. Due to the strong Hund's coupling, photoexcited down spin e$_{g\downarrow}$ electrons reverses the direction of t$_{2g}$ spins \cite{1998-PRB(R)_Matsuda}, which leads to AF coupling between neighbouring Mn ions. This spin disorder suppresses the DE mediated electron hopping, which relies on hopping between different Mn sites with parallel t$_{2g}$ spin alignment. Thus, photoinduced down spin optical transitions lead to suppression of ferromagnetic correlation between the spins and consequently deteriorates the metallic behaviour. Hence, an increase in resistance is observed on UV and blue illumination (Fig. 2(a)). Another important feature of the observed photoinduced effect is the persistent behaviour of PR (PPR) after photo illumination. PPR is more prominent after UV than blue illumination. In a previous report, Matsuda $\textit{et. al.}$ \cite{1998-PRB(R)_Matsuda} have studied the spin dynamics in (Nd$_{0.5}$Sm$_{0.5}$)$_{0.6}$Sr$_{0.4}$MnO$_3$ after e$_{g\downarrow}$ photoexcitation. They found that e$_{g\downarrow}$ excitation generates high frequency spin waves, that destroy spin correlation in the sample which leads to spin disorder. Their pump-probe experiments revealed that even after the relaxation of e$_{g\downarrow}$ spins ($\leq$1 ps), spin wave induced spin disorder persisted for a relatively long time ($\approx$300 ns) before the ferromagnetic ground state is recovered \cite{1998-PRB(R)_Matsuda}. In the present case, as resistivity measurement probe changes over a larger length scale (eg. magnetic domains) \cite{2000-JpnJAP-X_J_Liu}, the observed PPR may be a result of permanent change in the magnetic domains due to spin disorder. Also, the PPR effect shows that the observed PR is not due to thermal effects.
\begin{figure}
	\centering
	\includegraphics[width=1.0\linewidth]{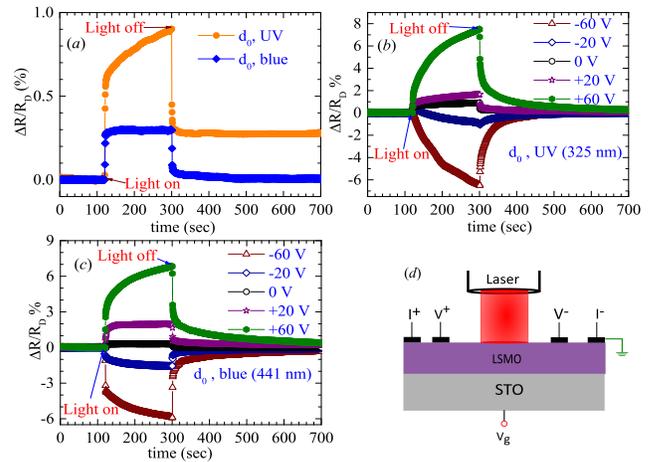}	
	\caption{(Color online) Plot of $\%$ resistance change of sample d$_0$  at 300 K. (a) For UV and blue illumination at V$_g$=0 (b) UV illumination at V$_g$$\neq$0 (c)  blue illumination at V$_g$$\neq$0. (d) Circuit configuration for measurement of the photoinduced effect.}
\end{figure}

A much more interesting physics is observed on applying a back gate voltage (V$_g$) to the samples in concurrence with photo exposure. For a negative gate voltage electrons are pushed into the bulk of the film, away from the interface due to screening effects, whereas electrons accumulate at the film-substrate interface for a positive gate voltage \cite{2009-PRL-Dhoot, 2014-PRB-Rastogi}. The effect of light and gate voltage is summarized in Fig. 2 (b) and (c) for sample d$_0$. In comparison to V$_g$=0, the percentage resistivity change increases with V$_g$$\neq$0 (-6.4$\%$ for V$_g$=-60 and +7.5$\%$ for V$_g$=+60).  Most importantly, we observe an inversion of photoresistivity for negative V$_g$. We believe that the inverse photoresistivity (or photoconductivity (PC)) to be a result of electron injection into the LSMO film as a result of gate voltage and photoexcited electron generation in STO substrate. We rationalize our interpretation as follows. Firstly, penetration depth(1$/$absorption coefficient) in LSMO (estimated from the absorption coefficient obtained from Fig. 2(a) of reference \cite{2013-ThinSolidFilms-Cesaria}) is $\approx$ 645 $\AA$  and 1818 $\AA$ for UV and blue light, respectively. As the penetration depth exceeds the film thickness (150 $\AA$), the substrate (STO) states are also accessible for excitation to both UV and blue light. Secondly, STO has a band gap (E$_{gap}$) of 3.2 eV ($<$ E$_{uv}$=3.8 eV ) \cite{2000-APL-Katsu}, hence on exposure to UV, free electron-hole pairs are generated in STO. The negative gate voltage then serves to inject UV generated electrons from STO into the film. Since LSMO is half metallic with a spin polarization of 95 $\%$ \cite{2003_APL_M_Bowen}, the injected electrons are spin polarized; which leads to the observed drop in resistivity. On the other hand, for positive V$_g$, the negative screening charges blocks the injection of UV generated electrons into the film. However, holes generated act as traps for interfacial electrons on the LSMO side. As a result, number of electrons available for hopping in LSMO decreases and an increase in resistance is observed. Katsu $\textit{et. al.}$ have reported similar electron injection at La$_{0.8}$Sr$_{0.2}$MnO$_{3}$/STO interface on white light irradiation \cite{2000-APL-Katsu}.
For blue light illumination, V$_g$ induced changes in resistivity of sample d$_0$ exhibits similar behaviour to that of UV irradiation (Fig. 2 (c)). We, therefore envisage a similar mechanism at work here i.e. injection of photogenerated electrons for negative V$_g$ and trapping of conduction electrons at the interface for positive V$_g$. Since E$_{blue}$ $<$ E$_{gap}$, photogenerated electrons result from the excitation of in-gap states in STO. Photoexcitation of in-gap states (due to oxygen vacancy) in STO has been known to result in enhanced photoconductivity in oxide films \cite{2012-OptLett-Rastogi}. 
\begin{figure}
	\centering
	\includegraphics[width=1.0\linewidth]{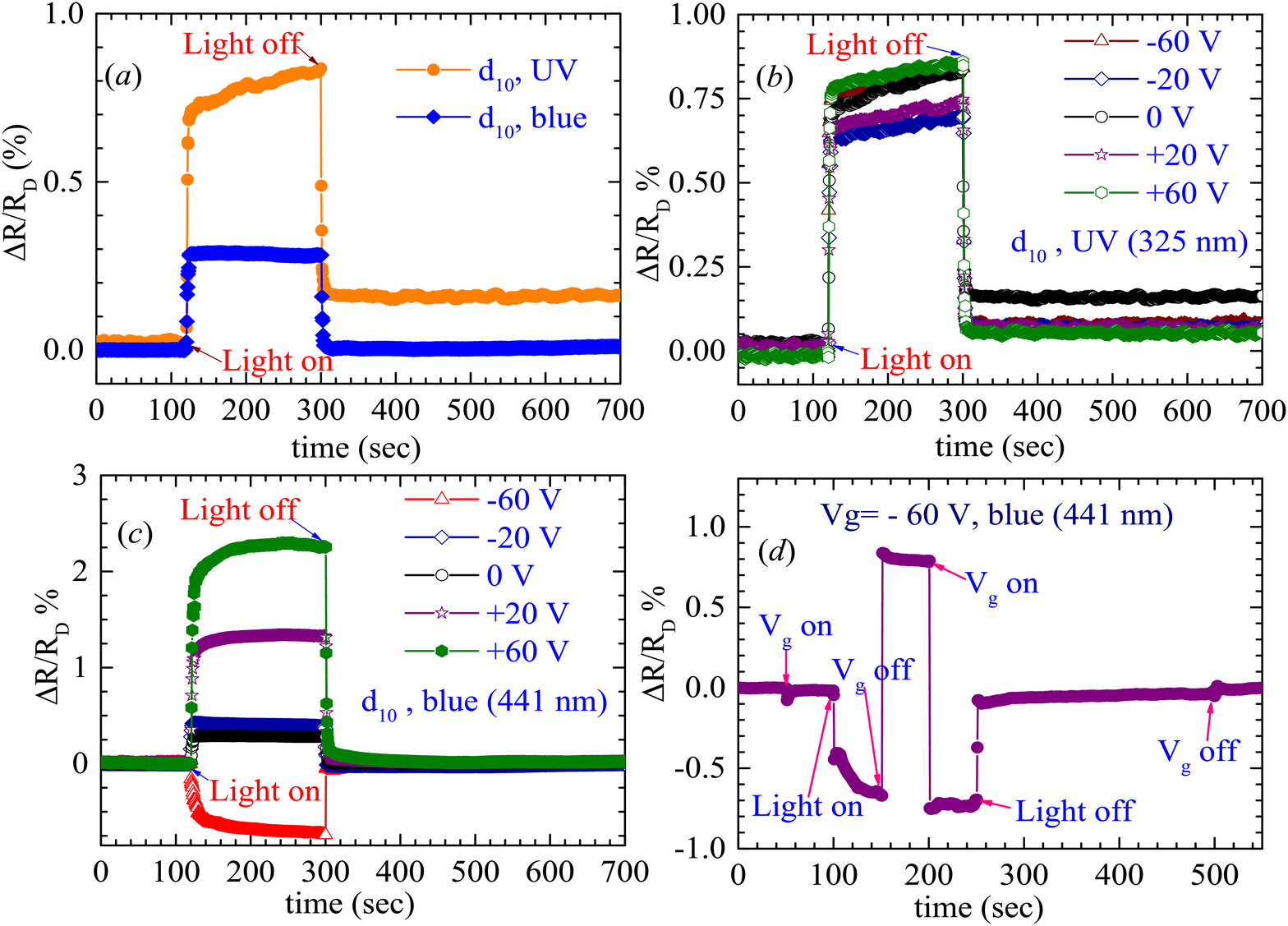}	
		\caption{(Color online) Plot of $\%$ resistance change of sample d$_{10}$  at 300 K. (a) For UV and blue illumination while V$_g$=0 (b) UV illumination while V$_g$$\neq$0 (c)  blue illumination while V$_g$$\neq$0 (d) resistance modulation of sample d$_{10}$ at V$_g$=-60 V and blue illumination.}
\end{figure}

To further verify the interfacial origin of the gating effect, we carry out photoinduced studies on sample d$_{10}$, which incorporates an insulating LTO layer at the interface. The results for UV and blue light are shown in Fig. 3. On UV irradiation, sample d$_{10}$ exhibits photoresistivity for V$_g$=0 as well as V$_g$$\neq$0 (Fig. 3(a)-(b)). Another important point to take note of is  the PR$\%$; which is almost of the same magnitude irrespective of the applied V$_g$ (PR$\%$$\approx$ +0.75$\%$) as shown in Fig. 3(b). Assuming carrier injection to be responsible for the observed photoinduced behaviour in sample d$_{0}$, the presence of insulating LTO layer at the interface should impede the carrier injection to some extent. For experiments carried out on sample  d$_{10}$ under UV illumination (Fig. 3(b)), PR$\%$ is independent of the applied V$_g$. Which in turn implies V$_g$ aided interfacial carrier injection is indeed blocked by the LTO layer. In the absence of gate voltage sample d$_{0}$ (without the interfacial LTO layer, Fig. 2(a)) and sample d$_{10}$ exhibit comparable value of PR$\%$ ($\approx$ +0.75$\%$) under UV illumination. This shows that the observed photoinduced effect in sample d$_{10}$ is entirely due to photoexcited e$_{g\downarrow}$ transitions due to UV illumination (even under applied V$_g$). This observation largely supports our conjecture of interfacial charge injection under applied V$_g$. And when the carrier injection is blocked in sample d$_{10}$, only PR from e$_{g\downarrow}$ transition is observed.

For blue light, sample d$_{10}$ (Fig. 3(a)) shows photoresistivity with comparable magnitude of PR$\%$ to sample d$_{0}$ (Fig. 2(a)) at V$_g$=0. This implies, blue light induced intersite Mn e$_{g\downarrow}$ transition is responsible for photoresistivity in sample d$_{10}$ at V$_g$=0. Whereas at V$_g$$\neq$0, a peculiar behaviour is observed (shown in Fig. 3 (c)), which signals a different mechanism for photoinduced effect during blue and UV illumination. At higher values of negative V$_g$(-60 V), photoconductivity persists. This observation indicates a different origin for the photoinduced behaviour other than charge injection for blue illumination. Fig. 3 (d) shows the modulation of resistance on application of V$_g$=-60 V and blue illumination for sample d$_{10}$. V$_g$ alone has negligible effect on the resistance. However, light exposure in presence of V$_g$  shows a substantial decrease in resistance. Furthermore, only light exposure (V$_g$ switched off) takes the sample to a resistive state. This clearly shows that photoconductivity is due to the  cooperative effect of light and V$_g$. 

Next, we study the effect of gate voltage on the photoinduced behaviour in LTO (10u.c)/STO film in order to discern the contribution of the interfacial LTO layer in sample d$_{10}$ (Fig. 4). LTO/STO interface is metallic owing to the presence of highly mobile interfacial 2D electron gas (2DEG), resulting from polar reconstruction at the interface \cite{2012-PRB-Rastogi,2018-JPCM-Ghising}. The 2DEG reside in a narrow potential well formed at the interface due to bending of STO conduction band. As shown in Fig. 4, the difference in photoinduced behaviour under applied V${_g}$ for UV and blue illumination indicates two separate mechanism for the two wavelengths. For UV illumination, photoconductivity is a direct result of generation and promotion of photoexcited electrons from valance to conduction band in STO, which contribute electrons to the 2DEG. For blue illumination, the appearance of PC at V$_g$=0, suggests the presence of in-gap states in STO (since E$_{blue}$ $<$ E$_{gap}$), which are excited by blue light. In LTO/STO the presence of oxygen vacancies (V$_O$) are responsible for the in-gap states. V$_O$ in STO are  generated due to the high gate voltages and during the film deposition process itself \cite{2007-PRB(R)-Kalabukhov, 2017-AdvMater-Kang}. Moreover, LTO/STO films fabricated at high temperatures reportedly leads to V$_O$ formation due to large oxidation tendency of LTO \cite{2012-OptLett-Rastogi, 2012-PRB-Rastogi}. V$_O$ have been known to donate electrons to the 2DEG at the interface \cite{2014-NatComm-Lei}. It has been reported that these V$_O$ become highly mobile under the action of light and gate voltages when trapped electrons in the V$_O$ are photoexcited \cite{2014-NatComm-Lei, 2015-SciRep-Li}. Thus, for negative V$_g$, the positively charged V$_O$ migrate away from the interface towards the cathode (bottom of STO). Therefore, electron contribution to 2DEG from V$_O$ diminishes \cite{2014-NatComm-Lei}, leading to an increase in resistance for blue illumination. Whereas for positive V$_g$, V$_O$ migrate towards the interface, where it donates electrons to the 2DEG that leads to decrease in resistance. Thus, in presence of V$_g$ photoinduced behaviour originate from different mechanisms for the two wavelength. In the following paragraphs we will discuss the role of V$_O$ in the photoinduced behaviour of LSMO.
\begin{figure}[htp]
	\centering
	\includegraphics[width=1\linewidth]{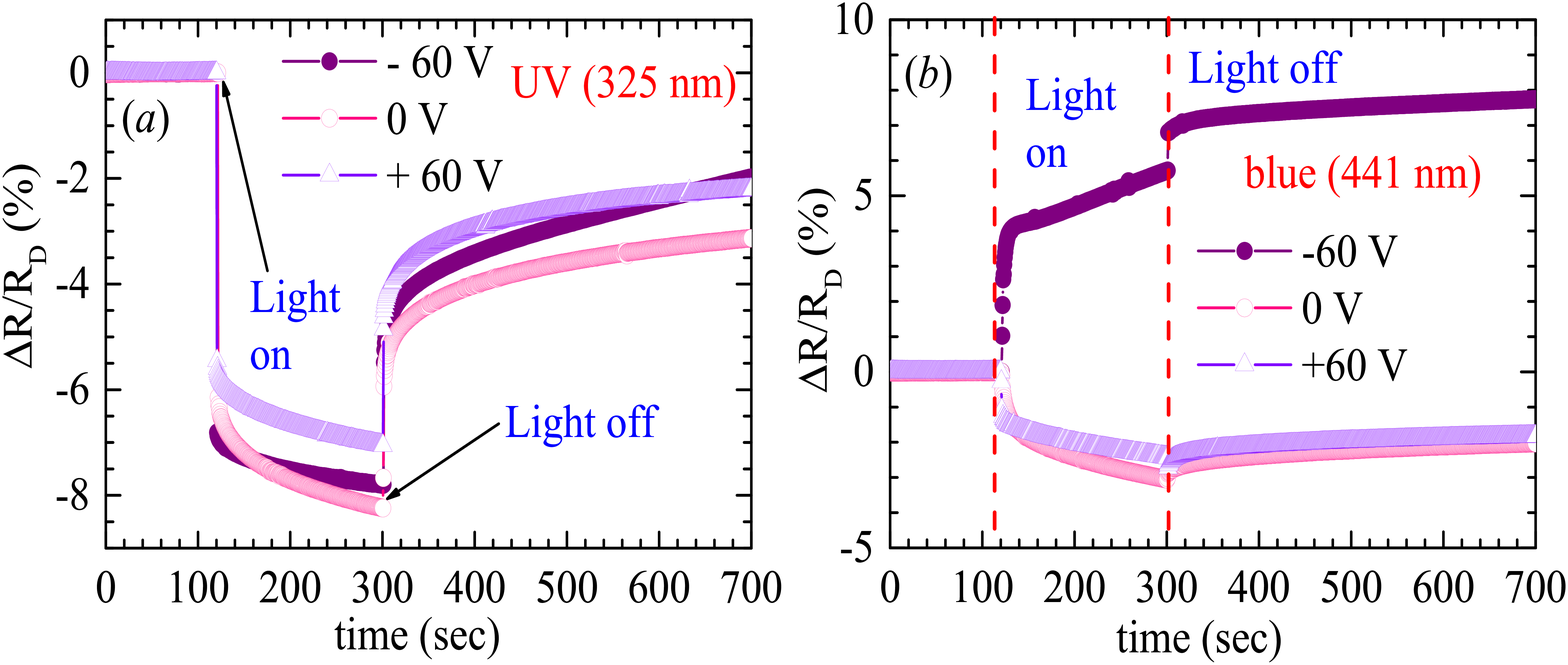}	
	\caption{(Color online)  Plot of $\%$ resistance change of LTO (10 u.c)/STO sample at different applied V$_g$ under (a) UV illumination and (b) blue illumination. }
\end{figure}

Since e$_g$ orbital occupancy in LSMO has tremendous influence on its magnetic and electrical properties, we need to look at the effect of V$_g$ and photo illumination on the e$_g$ orbital occupancy. In manganites tensile strain lowers the energy of d$_{x^2-y^2}$ orbital with respect to d$_{3z^2-r^2}$ orbital \cite{2008-PRB-Huijben,2006-PRB-Aruta}, which enhances its occupancy. Another factor which may also influence the orbital occupancy is V$_O$ in STO. At room temperature, V$_O$  in STO has been reported to be in the singly ionized state (V${_O^{\bullet}}$, i.e. V$_O$ donates one electron to the conduction band and retains the other electron) \cite{2007-PRL-Cuong,2013-PRL-Lin}. Photoexcitation of the in-gap states leaves V$_O$ in a doubly ionized V${_O^{\bullet\bullet}}$ state \cite{2015-SciRep-Li}. V${_O^{\bullet\bullet}}$ has a lower activation energy for diffusion(E$_A$=0.6 eV) as compared to V$_O^{\bullet}$ (E$_A$=1 eV) \cite{2015-SciRep-Li}; as a result V$_O$ diffusion process accelerates when photoexposure converts V${_O^{\bullet}}$ to V${_O^{\bullet\bullet}}$ \cite{2015-SciRep-Li}. In gated STO, V${_O}$ diffusion leads to accumulation of oxygen ions in the anode region and consequently to a a polar phase with elongated TiO$_6$ octahedra \cite{2013-PRB-Hanzig}. XRD measurements have confirmed an increase in out-of-plane lattice parameter (\textit{c}) in the anode region whereas it remains unchanged in the cathode region \cite{2013-PRB-Hanzig}. This result has important implication in our case; due to outward expansion of \textit{c} (along the z direction) at the STO interface resulting from V$_O$ diffusion, MnO$_6$ octahedra in the LSMO overlayers will be compressed along the z direction \cite{2014-NanoLett-Chen}. This is akin to a tensile strain in LSMO, which lowers d$_{x^2-y^2}$ orbital relative to d$_{3z^2-r^2}$.   Thus photoexposure and negative V$_g$ stabilizes d$_{x^2-y^2}$ orbitals. In films, in-plane correlations are much more important. Enhanced d$_{x^2-y^2}$ occupancy would boost planar exchange interaction, and so improvement in the ferromagnetic and metallic behaviour are to be expected \cite{2010-PRL-Sadoc}.

We would like to point out that the observed photoinduced effect in sample d$_{10}$ emanates from the LSMO film and not the LTO/STO or LSMO/LTO interface. As can be seen in Fig 3(b)-(c) and 4 (a)-(b) the photoinduced behaviour in LTO/STO sample is in contrast to that observed in sample d$_{10}$, so we can rule out the contribution of LTO/STO interface. Next we look at the LSMO/LTO interface. This interface also presents the possibility of polar discontinuity and charge transfer which could result in a photoinduced behaviour similar to LTO/STO sample. In the absence of V$_O$ in LTO, the photoinduced behaviour of the LSMO/LTO should be wavelength independent. This is because, the band gap of LTO (E$_{LTO}$ = 0.2 eV) is less than E$_{uv}$ (3.8 eV) and E$_{blue}$ (2.8 eV), so both UV and blue light can excite electrons from LTO valence band to conduction band; hence a similar trend in photoinduced behaviour would be seen for both wavelengths. Whereas in the case of STO, E$_{blue}$$<$E$_{STO}$$<$E$_{uv}$ and consequently photoinduced behaviour is different for the two wavelengths as explained earlier. As the photoinduced behaviour for blue and UV illumination is different (Fig. 3(b) and (c)) in sample d$_{10}$, we can exclude any contribution from LSMO/LTO interface to the observed photoinduced effect. Furthermore, if one considers the possibility of polar discontinuity and charge transfer, the LSMO layers have a charge of $\pm$0.7 e, while the LTO layers have a charge of $\pm$1 e. Therefore, polar discontinuity dictates electron transfer at the LSMO/LTO interface. However, studies on a similar heterostructure (LaAlO$_3$/La$_{0.67}$Sr$_{0.33}$MnO$_3$) have shown that no charge transfer occurs at such an interface \cite{2015-NatMat-Chen}. This result indicates that the polar discontinuity is not strong enough to induce a charge transfer and the lattice compensates for the polar discontinuity at such an interface by some other means  i.e. structural distortions, screening of polar field by the mobile electrons in metallic LSMO, etc.  A recent work by Saghayezhian \textit{et.al.} \cite{2019-PNAS-Saghayezhian}  have also shown that the charge transfer on account of polar discontinuity at the metal (LSMO)-insulator (STO) interface does not take place.

As we have already established the role of V$_O$ diffusion in Fig. 4(b), it plays a dominant role in the observed light induced resistivity change in sample d$_{10}$. When illuminated with UV (E$_{uv}$ $>$ E$_{gap}$), carrier generation in STO involves excitation of electrons from the valence band to the conduction band, while very less or no in-gap states are excited. However, in the case of blue light (E$_{blue}$ $<$ E$_{gap}$), only in-gap states in STO are excited which converts V${_O^{\bullet}}$ to highly mobile V${_O^{\bullet\bullet}}$ as explained in ref \cite{2015-SciRep-Li}. Thus, during blue illumination large number of highly mobile V${_O^{\bullet\bullet}}$ are generated in the STO substrate as compared to UV illumination. Since V$_O$ move due to the motion of oxygen ions in the opposite direction, they have the same mobility. During blue illumination, the large number of V${_O^{\bullet\bullet}}$ easily move towards the negative V$_g$ electrode. In response to this, oxygen ions in STO move in the opposite direction and reach the LSMO/LTO interface, diffusing through the LTO layer. The diffusion of oxygen ions through the insulating LTO layer may be facilitated by the similar structure of TiO$_6$ octahedra in LTO and STO. Drawing inspiration from the results of Hanzig \textit{et.al.} \cite{2013-PRB-Hanzig}, accumulation of oxygen ions at the LSMO/LTO interface would polarize the LTO layer resulting in an elongated TiO$_6$ octahedra perpendicular to the interface (towards the LSMO overlayer). This in turn would increase \textit{c} of the LTO layer. Consequently, MnO$_6$ octahedra in the LSMO overlayers will be compressed along the \textit{z} direction (as the in-plane lattice parameters of the LSMO overlayer are constrained by the STO substrate) \cite{2014-NanoLett-Chen}. This stabilizes d$_{x^2-y^2}$ in LSMO, that leads to an increase in in-plane hopping and consequently to a decrease in resistance \cite{2014-NanoLett-Chen, 2016-AdvFuncMater-Liu}. The observed photoconductivity at V$_g$=-60 V is due to stabilization of d$_{x^2-y^2}$, which increases electron hopping along the in-plane direction and increases the conductivity.  At a lower V$_g$ (-20 V), V$_O$ diffusion is subdued and the strain effect on MnO$_6$ octahedra vanish. As a result, photoresistivity is dominated by intersite e$_{g\downarrow}$ transition (since PR$\%$ is comparable to that in Fig. 3(a)). For positive V$_g$, V${_O^{\bullet\bullet}}$ travel to the LSMO/LTO interface which act as traps for electrons hopping into V${_O^{\bullet\bullet}}$ sites from the LSMO side. Similar studies of electron hopping across the interface and into the V$_O$ sites have been reported in oxide films \cite{2014-AMI-Wu}. This leads to decrease in conduction electrons in the LSMO overlayers and PR$\%$ increases with positive V$_g$. 

\begin{figure}
	\centering
	\includegraphics[width=1.0\linewidth]{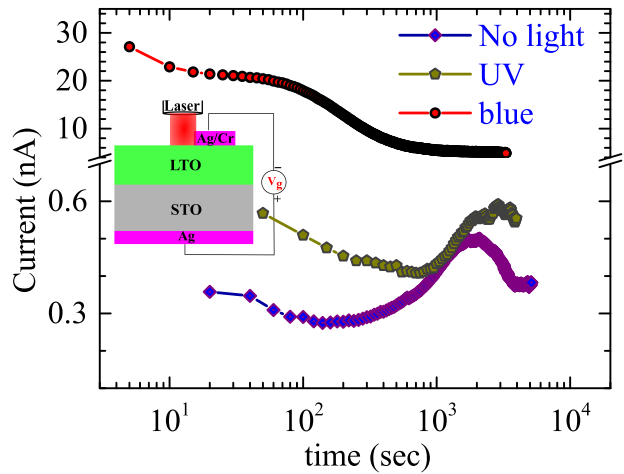}	
	\caption{(Color online) Transient current measurement of LTO(10 u.c)/STO sample at V$_g$=-60 V. The measurements were performed under dark (no illumination), UV and blue illumination. Inset shows the measurement configuration for transient current.}
\end{figure}
In order to study the dynamics of V$_O$ in our system, we have measured transient current in the presence of light and gate voltage. The circuit configuration for current measurement is shown in the inset of Fig. 5. A single step voltage is applied and temporal evolution of current is studied. In insulating titanates, oxygen vacancies are highly mobile \cite{1998_APL_S_Zafar}. Thus, the observed current is attributed to the diffusion of V$_O$ in the presence of applied voltage. According to space-charge-limited (SCL) theory, the transient current exhibits a peak, which is directly related to the mobility of the injected charge species \cite{1998_APL_S_Zafar}. Fig. 5 shows current as a function of time after a single step voltage (V$_g$=-60 V) is applied to the bottom electrode. The peak in the I(t) curve determines the mobility of V$_O$ during and in absence of illumination. According to SCL theory, the peak position ($\tau_p$) is related to the mobility ($\mu$) as follows \cite{1998_APL_S_Zafar}:
$$\tau_p=\frac{0.78d^2}{\mu V_g}$$
where \textit{d} is the sample thickness. The value of $\mu$ can be estimated from the above equation; the calculated values of $\mu$ are $5.0\times 10^{-7}$ cm$^2$/Vs for blue illumination, $1.5\times 10^{-8}$ cm$^2$/Vs for UV illumination and $1.7\times 10^{-8}$ cm$^2$/Vs for no light. As can be seen, while $\mu$ is of the same order for UV and for no light, it increases by an order of magnitude for blue light. Thus for blue light, generation of highly mobile V${_O^{\bullet\bullet}}$ due to photoexcitation of in-gap states leads to the observed photoconductivity in sample d$_{10}$. On the other hand,  UV illumination generates a large number of mobile electrons in STO rather than V${_O^{\bullet\bullet}}$, whose injection into the LSMO layer is blocked by the insulating LTO layer. As a result, UV illumination produces no changes in photoinduced behaviour with applied V$_g$ in sample d$_{10}$. And finally, now that we have established the role of V${_O^{\bullet\bullet}}$ in the photoinduced effect, it is reasonable to suggest that the photoinduced effect of sample d$_{0}$ (under blue illumination) originates from V${_O^{\bullet\bullet}}$ diffusion and electron injection.

In summary, we have studied the photoinduced effect on LSMO films at 300 K. Under UV and blue light exposure, depending on the sign of the applied gate voltage, both photoresistivity and photoconductivity are observed. While at V$_g$=0, spin disorder in LSMO induced by the e$_{g\downarrow}$ photoexcited transition is the dominant mechanism for photoresistivity, for V$_g$$\neq$0 interfacial charge injection and oxygen vacancy diffusion play a pivotal role in the photoinduced behaviour. Under UV illumination, photoinduced behaviour at V$_g$$\neq$0 stems from the injection of photoexcited electrons generated in the STO substrate. As a result, introducing an ultra thin layer of LTO at the LSMO/STO interface inhibits this carrier injection process and only photoresistivity from photoexcited e$_{g\downarrow}$ transition is observed (even for V$_g$$\neq$0). However, for blue illumination, the observation of photoconductivity (at high -V$_g$) even in the presence of interfacial LTO layer suggests the role of oxygen vacancy diffusion in the observed photoinduced effect at V$_g$$\neq$0. Light and gate voltage driven diffusion of oxygen vacancy in these heterostructure alters the orbital occupancy of e$_{g}$ electrons in the LSMO overlayer. In this work we have shown that using light and electric field one can simultaneously control the orbital and charge degrees of freedom in LSMO/LTO/STO heterostructure.

The authors would like to thank R. C. Budhani and S. Middey for the valuable discussions. Financial support from IIT Kanpur is gratefully acknowledged.

\bibliography{Reference}
\end{document}